\documentclass[preprint,eqsecnum,preprnumbers,nofootinbib,byrevtex,prd,aps,showpacs,showkeys,groupedaddress,floatfix]{revtex4}
\usepackage{bm}
\usepackage{graphics}
\usepackage{graphicx}
\usepackage{epsfig}
\usepackage{amssymb}
\usepackage{amsmath}
\begin{document}
\title{MESON PRODUCTION IN PROTON-PROTON COLLISIONS IN THE NAIVE NON-ABELIANIZATION APPROXIMATION  AND THE ROLE OF INFRARED RENORMALONS}
\author{A.~I.~Ahmadov$^{1,2}$~\footnote{ahmadovazar@yahoo.com}}
\author{Sh.~M.~Nagiyev$^{3}$}
\author{E.~A.~Dadashov$^{3}$}
\affiliation{$^{1}$ Department of Theoretical Physics, Baku State
University  \\ Z. Khalilov Street 23, AZ-1148, Baku, Azerbaijan}
\affiliation{$^{2}$ TH Division, Physics Department, CERN\\
CH-1211 Geneva 23, Switzerland}
\affiliation{$^{3}$Institute of
Physics of Azerbaijan National Academy of Sciences, H. Javid
Avenue, 33, AZ-1143, Baku, Azerbaijan}

\date{\today}

\begin{abstract}We calculate the "naive
non-abelianization" (NNA) contributions of the higher-twist Feynman
diagrams to the large-$p_T$  inclusive pion production cross section
in proton-proton collisions in the case of the running coupling and
frozen coupling approaches. We compare the resummed "naive
non-abelianization"  higher-twist cross sections with the ones
obtained in the framework of the frozen coupling approach and
leading-twist cross section. The structure of infrared renormalon
singularities of the higher twist subprocess cross section and it's
resummed expression are found. We discuss the phenomenological
consequences of possible higher-twist contributions to the pion
production in proton-proton collisions in within NNA.
\end{abstract}

\maketitle

\section{\bf Introduction}
The hadronic wave functions in terms of quark and gluon degrees of
freedom play an important role in the quantum chromodynamics
predictions for hadronic processes. If the hadronic wave functions
were accurately known, then we could calculate the hadronic
distribution amplitude and structure functions for exclusive and
inclusive processes in quantum chromodynamics (QCD).

The large-order behavior of a perturbative expansion in gauge
theories is inevitably dominated by the factorial growth of
renormalon diagrams [1-4]. In the case of QCD, the coefficients of
perturbative expansions in the QCD coupling $\alpha_{s}$ can
increase dramatically even at low orders. This fact, together with
the apparent freedom in the choice of renormalization scheme and
renormalization scales, limits the predictive power of
perturbative calculations, even in applications involving large
momentum transfers, where $\alpha_{s}$ is effectively small.
Investigation of the infrared renormalon effects in various
inclusive and exclusive processes is one of the most important and
interesting problem in the perturbative QCD (pQCD).

The frozen coupling constant approach in Refs.[5,6,7,8,9] was used
for calculation of integrals, such as
\begin{equation}
I\sim \int\frac{\alpha_{s}({Q}^2)\Phi(x,{Q}^2)}{1-x}dx
\end{equation}
It should be noted that, in pQCD calculations, the argument
${Q}^2$ of the running coupling constant  should be taken equal to
the square of the momentum transfer of a hard gluon in a
corresponding Feynman diagram, in both the renormalization and
factorization scale. But definýng in this way, $\alpha_{s}({Q}^2)$
suffers from infrared singularities. Therefore, in the soft
regions as $x_{1}\rightarrow 0$, and $x_{2}\rightarrow 0$, the
integral in (1.1) diverge and we need some regularization methods
for $\alpha_{s}(Q^2)$ in these regions for their calculation. It
is known that infrared renormalons are responsible for factorial
growth of coefficients in perturbative series for the physical
quantities. But, these divergent  series can be resummed by means
of the Borel transformation [1] and the principal  value
prescription [10], and effects of infrared renormalons can be
taken into account by a scale-setting procedure
$\alpha_{s}(Q^2)\rightarrow\alpha_{s}(exp(f(Q^2))Q^2)$ at the
one-loop order results.

In this work we apply the running coupling approach [11] in order
to compute the effects of the infrared renormalons on the meson
production in proton-proton collisions in "naive
non-abelianization" approximation. This approach was also employed
previously [12-14] to calculate the inclusive meson production in
proton-proton and photon-photon collisions. The running coupling
approach in "naive non-abelianization" approximation  pion
electromagnetic form factor was computed in [15].

A precise measurement of the inclusive charged pion production cross
section at $\sqrt s = 62.4\,\,GeV$ and $\sqrt s = 200\,\,GeV$ is
important for the proton-proton collisions program at  the
Relativistic Heavy Ion Collider (RHIC) at the Brookhaven National
Laboratory.

Therefore, it will be interesting that the calculation of the
higher-twist effects on the dependence of the pion wave function in
pion production at proton-proton collisions by the running coupling
constant approach. In this respect, the contribution of the
higher-twist Feynman diagrams to a pion production cross section in
proton-proton collisions has been computed by using the the infrared
renormalon improved distribution amplitude of the pion. Also,
higher-twist contributions which are calculated by the running
coupling constant and frozen coupling constant approaches are been
estimated and compared to each other. Within this context, this
paper is organized as follows: In Sec.~\ref{ht}, we provide formulas
for the calculation of the contribution of the higher twist and
leading twist diagrams. In Sec.~\ref{ir} we present formulas and an
analysis of the higher-twist effects on the dependence of the pion
wave function by the running coupling constant approach. In
Sec.~\ref{results}, we give the numerical results for the cross
section and discuss the dependence of the cross section on the pion
wave functions. We present our conclusions in Sec.~\ref{conc}.

\section{THE HIGHER TWIST AND LEADING TWIST CONTRIBUTIONS TO INCLUSIVE REACTIONS}
\label{ht} The higher-twist Feynman diagrams, which describe the
subprocess $q_1+\bar{q}_{2} \to \pi^{+}(\pi^{-})+\gamma$ for the
pion production in the proton-proton collision are shown in Fig.1.
The amplitude for this subprocess can be found by means of the
Brodsky-Lepage formula [16]:
\begin{equation}
M(\hat s,\hat
t)=\int_{0}^{1}{dx_1}\int_{0}^{1}dx_2\delta(1-x_1-x_2)\Phi_{\pi}(x_1,x_2,Q^2)T_{H}(\hat
s,\hat t;x_1,x_2).
\end{equation}
In Eq.(2.1), $T_H$ is  the sum of the graphs contributing to the
hard-scattering part of the subprocess. In our calculation, we have
neglected the pion and the proton masses. The Mandelstam invariant
variables for subprocesses $q_1+\bar{q}_{2} \to
\pi^{+}(\pi^{-})+\gamma$ are defined as
\begin{equation}
\hat s=(p_1+p_2)^2,\quad \hat t=(p_1-p_{\pi})^2,\quad \hat
u=(p_1-p_{\gamma})^2.
\end{equation}

One of the important moment in our study is the choice of the pion
wave functions $\Phi_{\pi}(x,Q^2)$ in Eq.(2.1). In Ref.[16] the
authors have calculated the contribution of "bubble chain" diagrams
to the Brodsky-Lepage evolution kernel $V[x,y;\alpha(Q^2)]$ in the
"naive non-abelianization" (NNA) approximation and, as a result,
have got new, infrared renormalon improved distribution amplitude
for the meson in the form:
\begin{equation}
\Phi_{\pi}(x,Q^2)=f_{M}[x(1-x)]^{1+\alpha}\sum_{n=0}^{\infty}a_{n}(Q^2)A_{n}C_{n}^{3/2+\alpha}(2x-1)
\end{equation}
where ${C_{n}^{3/2+\alpha}(2x-1)}$  are the Gegenbauer
polynomials, $A_{n}(\alpha_{s})$ are normalization constants,
$a_{n}(Q^2)$ define the evolution of $\Phi_{\pi}(x,Q^2)$ with
$Q^2$ and $\alpha\equiv\alpha_{s}(Q^2)\beta_{0}/4\pi$. In our
calculations are normalization constants $A_{n}(\alpha_{s})$ are
given by the expression
\begin{equation}
A_{n}(\alpha_{s})=\frac{\Gamma(3+2\alpha)}{\sqrt{3}\Gamma(1+\alpha)\Gamma(2+\alpha)}\cdot\frac{n!}{(2+2\alpha)_{n}}\cdot\frac{3+2\alpha+2n}{2+2\alpha+n}.
\end{equation}
where $\Gamma(a)$ is the Euler gamma function, $(a)_n$ is the
Pochhammer symbol, $(a)_n=\Gamma(a+n)/\Gamma(a)$. It should be
noted that normalization constants $A_{n}(\alpha_{s})$ given by
expression (2.4) differ from ones of Ref.[17]. If in the Ref.[17]
change $\alpha$ by $\alpha+1$, then we obtain expression (2.4). We
have used three different wave functions: the asymptotic  (asy),
and the Gosdzinsky-Kivel wave functions [17].
$$
\Phi_{GK1}(x,\mu_{0}^2)=\Phi_{asy}(x,\alpha)\left[C_{0}^{3/2+\alpha}(2x-1)+1.7A_{2}C_{2}^{3/2+\alpha}(2x-1)\right],
$$
$$
\Phi_{GK2}(x,\mu_{0}^2)=\Phi_{asy}(x,\alpha)\left[C_{0}^{3/2+\alpha}(2x-1)+1.7A_{2}C_{2}^{3/2+\alpha}(2x-1)+1.6A_{4}C_{4}^{3/2+\alpha}(2x-1)\right],$$
\begin{equation}
\Phi_{asy}(x,\alpha)=f_{\pi}\frac{\Gamma(4+2\alpha)}{2\Gamma^2(2+\alpha)}[x(1-x)]^{1+\alpha},
\end{equation}
$$
C_{0}^{3/2+\alpha}(2x-1)=1,\,\,C_{2}^{3/2+\alpha}(2x-1)=\frac{3+2\alpha}{2}((5+2\alpha)(2x-1)^2-1),
$$
$$
C_{4}^{3/2+\alpha}(2x-1)=\frac{(3+2\alpha)(5+2\alpha)}{24}[(9+2\alpha)(7+2\alpha)(2x-1)^4-6(7+2\alpha)(2x-1)^2+3]
$$
where $f_{\pi}$ is the pion decay constant. The evolution of the
wave function on the factorization scale $Q^2$ is governed by the
functions $a_n(Q^2)$,
\begin {equation}
a_n(Q^2)=a_n(\mu_{0}^2)exp\left[\int^{\alpha_{s}(p_{T}^2)}_{\alpha_{s}(\mu_{0}^2)}4\pi\frac{\gamma_{k}(x)+C_{kk}(x)}{\beta_{0}x^2}dx\right],
\end{equation}
where
\begin {equation}
C_{kk}(\alpha_{s})=(\alpha_{s}\frac{\beta_{0}}{4\pi})^2\{\psi(\frac{3}{2}+\alpha+2k)-\psi(\frac{3}{2}+\alpha)\},
\end{equation}
\begin {equation}
\gamma_{n}=C_{F}\frac{(1+\alpha)^2\Gamma(2\alpha+4)}{3(2+\alpha)\Gamma(2+\alpha)^3\Gamma(1-\alpha)}\{1-\frac{(\alpha+1)(\alpha+2)}{(\alpha+1+n)(\alpha+n+2)}+\frac{2(\alpha+2)}{\alpha+1}[\psi(\alpha+2+n)-\psi(\alpha+2)]\}.
\end{equation}
In the limit $\alpha\rightarrow 0$ from infrared renormalon improved
distribution amplitude we can obtain ordinary distribution
amplitude.

The cross section for the higher-twist subprocess $q_1\bar{q}_{2}
\to \pi^{+}(\pi^{-})\gamma$ is given by the expression
\begin{equation}
\frac{d\sigma}{d\hat t}(\hat s,\hat t,\hat u)=\frac
{8\pi^2\alpha_{E} C_F}{27}\frac{\left[D(\hat t,\hat
u)\right]^2}{{\hat s}^3}\left[\frac{1}{{\hat u}^2}+\frac{1}{{\hat
t}^2}\right],
\end{equation}
where
\begin{equation}
D(\hat t,\hat u)=e_1\hat
t\int_{0}^{1}dx\left[\frac{\alpha_{s}(Q_1^2)\Phi_{\pi}(x,
Q_1^2)}{1-x}\right]+e_2\hat u\int_{0}^{1}dx\left[\frac{\alpha_{s}(
Q_2^2)\Phi_{\pi}(x,Q_2^2)}{1-x}\right].
\end{equation}
Here $Q_{1}^2=(x-1)\hat u,\,\,\,\,$and $Q_{2}^2=-x\hat t$,\,\,
represent the momentum squared carried by the hard gluon in Fig.1,
$e_1(e_2)$ is the charge of $q_1(\overline{q}_2)$ and
$C_F=\frac{4}{3}$. The higher-twist contribution to the
large-$p_{T}$ pion production cross section in the process
$pp\to\pi^{+}(\pi^{-})+\gamma+X$ is [18,19]:
\begin{equation}
\Sigma_{M}^{HT}\equiv E\frac{d\sigma}{d^3p}=\int_{0}^{1}\int_{0}^{1}
dx_1 dx_2 G_{{q_{1}}/{h_{1}}}(x_{1})
G_{{q_{2}}/{h_{2}}}(x_{2})\frac{\hat s}{\pi} \frac{d\sigma}{d\hat
t}(q\overline{q}\to \pi\gamma)\delta(\hat s+\hat t+\hat u).
\end{equation}

In the numerical calculations we denote the higher-twist cross
section obtained using the frozen coupling constant approximation by
$(\Sigma_{\pi}^{HT})^0$.

One of the essential problem, extracting the higher-twist
corrections to the pion production cross section and a comparison of
higher-twist corrections with leading-twist contributions. We take
two leading-twist subprocesses for the pion production:(1)
quark-antiquark annihilation $q\bar{q} \to g\gamma$, in which  $g
\to \pi^{+}(\pi^{-})$ and (2) quark-gluon fusion, $qg \to q\gamma $,
with subsequent fragmentation of the final quark into a meson, $q
\to \pi^{+}(\pi^{-})$.

\section{HIGHER-TWIST MECHANISM AND THE ROLE OF INFRARED RENORMALONS}\label{ir}
In this section, we will calculate the integral (2.10) using the
running coupling constant approach in the naive non-abelianization
approximation and also discuss the problem of normalization of the
higher-twist process cross section in the context of the same
approach. As  is seen from (2.10), in general, one has to take
into account not only the dependence of $\alpha( {Q}^2)$ on the
scale ${Q}^2$, but also an evolution of $\Phi(x,{Q}^2)$ with
${Q}^2$. Therefore, it is worth noting that, the renormalization
scale (argument of $\alpha_s$) should be equal to
$Q_{1}^2=(x-1)\hat u$, $Q_{2}^2=-x\hat t$, whereas the
factorization scale [$Q^2$ in $\Phi_{M}(x,Q^2)$] is taken
independent from $x$, we assume $Q^2=p_{T}^2$. It should be noted
in Ref.[17] the authors also noted the existence of two kinds of
power corrections to cross sections; the infrared renormalon
ambiguity arising from the loop integration and power corrections
from regions as, $x\rightarrow0$, $x\rightarrow1$.  The integral
(2.10) in the framework of the running coupling approach takes the
form
\begin{equation}
I(\mu_{R_{0}}^2)=\int_{0}^{1}\frac{\alpha_{s}(\lambda
\mu_{R_0}^2)\Phi_{M}(x,\mu_{F}^2)dx}{1-x}.
\end{equation}
The $\alpha_{s}(\lambda \mu_{R_0}^2)$ has the infrared singularity
at $x\rightarrow1$, for $\lambda=1-x$  or $x\rightarrow0$, for
$\lambda=x$ and so the integral $(3.1)$ diverges. Hence, the
integral (3.1) can be found after regularization of
$\alpha_{s}(\lambda \mu_{R_0}^2)$ in these end point regions. Such
regularization can be fulfilled with the aid of the
renormalization group equation that allows us to express the
running coupling constant $\alpha_{s}(\lambda Q^2)$ in terms of
$\alpha_{s}(Q^2)$. The solution of renormalization group equation
for the running coupling $\alpha\equiv\alpha_{s}/\pi$ has the form
[10]
\begin{equation}
\frac{\alpha(\lambda)}{\alpha}=\left[1+\alpha
\frac{\beta_{0}}{4}\ln{\lambda}\right]^{-1}.
\end{equation}
Then for $\alpha(\lambda Q^2)$, we get
\begin{equation}
\alpha(\lambda Q^2)=\frac{\alpha_{s}}{1+\ln{\lambda/t}}.
\end{equation}
where $t=4\pi/\alpha_{s}(Q^2)\beta_{0}=4/\alpha\beta_{0}$.

Having inserted  Eq.(3.3) into Eq.(2.10) we obtain
$$
D(\hat t,\hat u)= e_{1}\hat t\alpha_{s}(-\hat u)t_{1}\int_{0}^{1}dx
\frac{f_{M}[x(1-x)]^{1+\alpha}\sum_{n=0}^{\infty}a_{n}(Q^2)A_{n}C_{n}^{3/2+\alpha}(2x-1)}{(1-x)
(t_{1}+\ln\lambda)}+
$$
\begin{equation}
e_{2}\hat u\alpha_{s}(-\hat t)t_{2}\int_{0}^{1}dx
\frac{f_{M}[x(1-x)]^{1+\alpha}\sum_{n=0}^{\infty}a_{n}(Q^2)A_{n}C_{n}^{3/2+\alpha}(2x-1)}{(1-x)
(t_{2}+\ln\lambda)}
\end{equation}

where $t_1=4\pi/\alpha_{s}(-\hat u)\beta_{0}$ and
$t_2=4\pi/\alpha_{s}(-\hat t)\beta_{0}$.

The integral (3.4) as we know, still divergent, which  can be
defined by existing methods. Using the running coupling approach it
may be found as a perturbative series in $\alpha_{s}(Q^2)$ with
factorially growing coefficient. Making the change variable as
$z=\ln\lambda$ we obtain
\begin{equation}
D(\hat t,\hat u)=e_{1}\hat t \alpha_{s}(-\hat u) t_1\int_{0}^{1}dx
\frac{\Phi_{M}(x,p_{T}^2)}{(1-x)(t_1+z)}+ e_{2}\hat u
\alpha_{s}(-\hat t) t_2 \int_{0}^{1} dx
\frac{\Phi_{M}(x,p_{T}^2)}{(1-x)(t_2+z)}
\end{equation}
In order to calculate (3.5) we will apply the integral
representation of $1/(t+z)$ [20,21].
\begin{equation}
\frac{1}{(t+z)}=\int_{0}^{\infty}e^{-(t+z)u}du
\end{equation}
gives
$$
D(\hat t,\hat u)=e_{1} \hat{t} \alpha_{s}(-\hat u) t_1
\int_{0}^{1} \int_{0}^{\infty}
\frac{\Phi_{M}(x,p_{T}^2)e^{-(t_1+z)u}du dx}{(1-x)}+
$$
\begin{equation}
e_{2} \hat{u} \alpha_{s}(-\hat t) t_2 \int_{0}^{1} \int_{0}^{\infty}
\frac{\Phi_{M}(x,p_{T}^2)e^{-(t_2+z)u}du dx}{(1-x)}.
\end{equation}

In the case  $\Phi_{asy}(x,\alpha)$ for the $D(\hat t,\hat u)$, it
is written as
$$
D(\hat t,\hat u)=\frac{4\pi f_{\pi} e_{1} \hat
t}{\beta_{0}}\cdot\frac{\Gamma(4+2\alpha)}{2\sqrt{3}\Gamma^2(2+\alpha)}\cdot
\int_{0}^{\infty} du e^{-t_{1}u}B(2+\alpha,1+\alpha-u)+
$$
\begin{equation}
\frac{4\pi f_{\pi} e_{2} \hat
u}{\beta_{0}}\cdot\frac{\Gamma(4+2\alpha)}{2\sqrt{3}\Gamma^2(2+\alpha)}\cdot
\int_{0}^{\infty} du e^{-t_{2}u}B(2+\alpha,1+\alpha-u).
\end{equation}
and for $\Phi_{GK1}(x,Q^2)$ wave function
$$
D(\hat t,\hat u)=\frac{4\pi f_{\pi} e_{1} \hat
t}{\beta_{0}}\cdot\frac{\Gamma(4+2\alpha)}{2\sqrt{3}\Gamma^2(2+\alpha)}\cdot
\int_{0}^{\infty} du
e^{-t_{1}u}[A_{0}(\alpha_{s})B(2+\alpha,1+\alpha-u)+1.7A_{2}(\alpha_{s})\frac{3+2\alpha}{2}\cdot
$$
$$
exp\left({\frac{4\pi}{\beta_{0}}}{\int}^{\alpha_{s}(p_{T}^2)}_{\alpha_{s}(\mu_{0}^2)}
\frac{\gamma_{k}(x)+C_{kk}(x)}{x^2}dx\right)((5+2\alpha)(4B(4+\alpha,1+\alpha-u)-
$$
$$
4B(3+\alpha,1+\alpha-u)+B(2+\alpha,1+\alpha-u))-B(2+\alpha,1+\alpha-u)]+
$$
$$
\frac{4\pi f_{\pi} e_{2} \hat
u}{\beta_{0}}\cdot\frac{\Gamma(4+2\alpha)}{2\sqrt{3}\Gamma^2(2+\alpha)}\cdot
\int_{0}^{\infty} du
e^{-t_{2}u}[A_{0}(\alpha_{s})B(2+\alpha,1+\alpha-u)+1.7A_{2}(\alpha_{s})\frac{3+2\alpha}{2}
$$
$$
exp\left({\frac{4\pi}{\beta_{0}}}{\int}^{\alpha_{s}(p_{T}^2)}_{\alpha_{s}(\mu_{0}^2)}\frac{\gamma_{k}(x)+C_{kk}(x)}{x^2}dx\right)((5+2\alpha)(4B(4+\alpha,1+\alpha-u)-
$$
\begin{equation}
4B(3+\alpha,1+\alpha-u)+B(2+\alpha,1+\alpha-u))-B(2+\alpha,1+\alpha-u)].
\end{equation}
also for $\Phi_{GK2}(x,Q^2)$ wave function
$$
D(\hat t,\hat u)=\frac{4\pi f_{\pi} e_{1} \hat
t}{\beta_{0}}\cdot\frac{\Gamma(4+2\alpha)}{2\sqrt{3}\Gamma^2(2+\alpha)}\cdot
\int_{0}^{\infty} du
e^{-t_{1}u}[A_{0}(\alpha_{s})B(2+\alpha,1+\alpha-u)+1.7A_{2}(\alpha_{s})\frac{3+2\alpha}{2}\cdot
$$
$$
exp\left({\frac{4\pi}{\beta_{0}}}{\int}^{\alpha_{s}(p_{T}^2)}_{\alpha_{s}(\mu_{0}^2)}
\frac{\gamma_{k}(x)+C_{kk}(x)}{x^2}dx\right)((5+2\alpha)(4B(4+\alpha,1+\alpha-u)-
$$
$$
4B(3+\alpha,1+\alpha-u)+B(2+\alpha,1+\alpha-u))-B(2+\alpha,1+\alpha-u)]+
$$
$$
1.6A_{4}(\alpha_{s})\frac{(3+2\alpha)(5+2\alpha)}{24}\cdot
$$
$$
exp\left({\frac{4\pi}{\beta_{0}}}{\int}^{\alpha_{s}(p_{T}^2)}_{\alpha_{s}(\mu_{0}^2)}
\frac{\gamma_{k}(x)+C_{kk}(x)}{x^2}dx\right)((9+2\alpha)(7+2\alpha)(16B(6+\alpha,1+\alpha-u)-
$$
$$
32B(5+\alpha,1+\alpha-u)+24B(4+\alpha,1+\alpha-u)-8B(3+\alpha,1+\alpha-u)+B(2+\alpha,1+\alpha-u))-
$$
$$
6(7+2\alpha)(4B(4+\alpha,1+\alpha-u)-
$$
$$
4B(3+\alpha,1+\alpha-u)+B(2+\alpha,1+\alpha-u))+3B(2+\alpha,1+\alpha-u)]+
$$
$$
\frac{4\pi f_{\pi} e_{2} \hat
u}{\beta_{0}}\cdot\frac{\Gamma(4+2\alpha)}{2\sqrt{3}\Gamma^2(2+\alpha)}\cdot
\int_{0}^{\infty} du
e^{-t_{2}u}[A_{0}(\alpha_{s})B(2+\alpha,1+\alpha-u)+1.7A_{2}(\alpha_{s})\frac{3+2\alpha}{2}\cdot
$$
$$
exp\left({\frac{4\pi}{\beta_{0}}}{\int}^{\alpha_{s}(p_{T}^2)}_{\alpha_{s}(\mu_{0}^2)}
\frac{\gamma_{k}(x)+C_{kk}(x)}{x^2}dx\right((5+2\alpha)(4B(4+\alpha,1+\alpha-u)-
$$
$$
4B(3+\alpha,1+\alpha-u)+B(2+\alpha,1+\alpha-u))-B(2+\alpha,1+\alpha-u)]+
$$
$$
1.6A_{4}(\alpha_{s})\frac{(3+2\alpha)(5+2\alpha)}{24}\cdot
$$
$$
exp\left({\frac{4\pi}{\beta_{0}}}{\int}^{\alpha_{s}(p_{T}^2)}_{\alpha_{s}(\mu_{0}^2)}
\frac{\gamma_{k}(x)+C_{kk}(x)}{x^2}dx\right)((9+2\alpha)(7+2\alpha)(16B(6+\alpha,1+\alpha-u)-
$$
$$
32B(5+\alpha,1+\alpha-u)+24B(4+\alpha,1+\alpha-u)-8B(3+\alpha,1+\alpha-u)+B(2+\alpha,1+\alpha-u))-
$$
$$
6(7+2\alpha)(4B(4+\alpha,1+\alpha-u)-
$$
\begin{equation}
4B(3+\alpha,1+\alpha-u)+B(2+\alpha,1+\alpha-u))+3B(2+\alpha,1+\alpha-u)].
\end{equation}
where $B(\alpha,\beta)$ is Beta function.

The inverse Borel transformations have the infinite number of
infrared renormalon poles at the points $u_{0} = k + \alpha$ in the
Borel plane. According to Ref.[4] that infrared renormalon pole at
$u_{0}=k$ correspond to a power-suppressed correction as
$(\Lambda^2/Q^2)^k$ to a physical quantity under consideration. If
the renormalon pole is located at $u_{0}=k+\alpha$ then its
contribution is order $(\Lambda^2/Q^2)^k/e$ or
$(\Lambda^2/Q^2)^k/e^2$ . Therefore, our expressions (3.8)-(3.10)
takes into account the power-suppressed corrections
$C_{k}(Q^2)(\Lambda^2/Q^2)^k$, $k=1,2,3,..$ to the inclusive pion
production cross section in proton-proton collisions. The
coefficients $C_{k}(Q^2)$ of these corrections also depend on the
chosen pion wave functions. It should be noted that here we neglect
infrared ambiguities $\delta C_{k}(Q^2)(\Lambda^2/Q^2)^k$ producing
by the principal value prescription itself, which have to be
canceled by ultraviolet renormalon ambiguities of higher twist
corrections to cross section and do not estimate $\delta
C_{k}(Q^2)$. The structure of the infrared renormalon poles in
Eqs.(3.8-3.10) strongly depends on the wave functions of the pion.
In the numerical calculations we denote resummed higher-twist cross
section by $(\Sigma_{\pi}^{HT})^{res}$.

\section{NUMERICAL RESULTS AND DISCUSSION}\label{results}

In this section,  the numerical results for the "naive
non-abelianization" contribution of higher-twist effects to
large-$p_{T}$ inclusive pion production cross section in the
process $pp \to \pi^{+}(or\,\, \pi^{-})\gamma +X$ in the case
higher-twist contributions calculated in the context of the
running coupling and frozen coupling approaches on the dependence
of the infrared renormalon improved distribution amplitude of the
pion are discussed. In the numerical calculations for the quark
distribution function inside the proton, the MSTW distribution
function  has been used [22]. The gluon and quark fragmentation
functions into a pion has been taken from [23]. The results of our
numerical calculations are plotted in Fig.2-Fig.9. In Fig.2 -
Fig.4 we show the dependence of the higher-twist cross sections
$(\Sigma_{\pi^{+}}^{HT})^{0}$, $(\Sigma_{\pi^{+}}^{HT})^{res}$
calculated in the context of the frozen and  running couplings
constant approaches and the ratios
$R=(\Sigma_{\pi^{+}}^{HT})^{res}$/$\Sigma_{\pi^{+}}^{HT})^{0}$,
$(\Sigma_{\pi^{+}}^{HT})^{0}$/$(\Sigma_{\pi^{+}}^{LT})$
$(\Sigma_{\pi^{+}}^{HT})^{res}$/$(\Sigma_{\pi^{+}}^{LT})$ as a
function of the pion transverse momentum $p_{T}$ for different
pion wave functions at $y=0$. It is seen from Fig.2 that the
higher-twist cross section is monotonically decreasing with an
increase in the transverse momentum of the pion. In Fig.3 - Fig.4,
we show the dependence of the ratios
$R=(\Sigma_{\pi^{+}}^{HT})^{res}$/$(\Sigma_{\pi^{+}}^{HT})^{0}$,
$(\Sigma_{\pi^{+}}^{HT})^{0}$/$\Sigma_{\pi^{+}}^{LT}$,
$(\Sigma_{\pi^{+}}^{HT})^{res}$/$\Sigma_{\pi^{+}}^{LT}$ as a
function of the pion transverse momentum $p_{T}$ for different
pion wave functions. Here  $\Sigma_{\pi^{+}}^{LT}$ is the
leading-twist cross section, respectively. As shown in Fig.3, in
the region $2\,\,GeV/c<p_T<4\,\,GeV/c$  resummed higher-twist
cross section is suppress by about 1-2 orders of magnitude
relative to the higher-twist cross section calculated in the
framework of the frozen coupling approach. In Fig.4, we show the
dependence of the ratios
$(\Sigma_{\pi^{+}}^{HT})^{0}$/$\Sigma_{\pi^{+}}^{LT}$, and
$(\Sigma_{\pi^{+}}^{HT})^{res}$/$\Sigma_{\pi^{+}}^{LT}$ as a
function of the pion transverse momentum $p_{T}$ for different
pion wave functions. It is observed from Fig.4 that, the ratio
$(\Sigma_{\pi^{+}}^{HT})^{0}$/$\Sigma_{\pi^{+}}^{LT}$ with
increasing transverse momentum of pion  decrease and has a minimum
approximately at the point $p_{T}=10GeV/c$. After that, the ratio
increase with increasing in the $p_{T}$ transverse momentum of the
pion. Also, as shown in Fig.4, the ratio
$(\Sigma_{\pi^{+}}^{HT})^{res}$/$\Sigma_{\pi^{+}}^{LT}$  decrease
with increasing in the $p_{T}$ transverse momentum of the pion. In
Fig.5, we have depicted higher-twist cross sections
$(\Sigma_{\pi^{+}}^{HT})^{0}$, $(\Sigma_{\pi^{+}}^{HT})^{res}$, as
a function of the rapidity $y$ of the pion at $\sqrt
s=62.4\,\,GeV$ and $p_T=4.9\,\,GeV/c$. Figure show that
higher-twist cross section in case frozen and running coupling
constant approaches have a different distinctive. In the region
($-2.52\leq y\leq 1.22$) the higher-twist cross section
$(\Sigma_{\pi^{+}}^{HT})^{0}$ for all wave functions increase with
an increase of the $y$ rapidity of the pion and has a maximum
approximately at the point $y=1.22$. But, the resummed
higher-twist cross section $(\Sigma_{\pi^{+}}^{HT})^{res}$ for all
wave functions increase with an increase of the $y$ rapidity of
the pion and has a maximum approximately at the point $y=-1.92$.
As is seen from Fig.5 in the region $-2.52\leq y\leq 2$ resummed
higher-twist cross section for $\Phi_{GK1}(x,Q^2)$ and
$\Phi_{GK2}(x,Q^2)$ is suppressed by about one order of magnitude
relative to the resummed higher-twist cross section for
$\Phi_{asy}(x,\alpha)$. As is seen from Fig.6 ratio
$R=(\Sigma_{\pi^{+}}^{HT})^{res}/(\Sigma_{\pi^{+}}^{HT})^{0}$,\,
for all wave functions increase with an increase of the $y$
rapidity of the pion and has a maximum approximately at the point
$y=-1.92$. In the region  $-2.52<y<-1.92$ resummed higher-twist
cross section is suppress by about two orders of magnitude
relative to the higher-twist cross section calculated in the
framework of the frozen coupling approach. Besides that, the ratio
decreases with an increase in the $y$ rapidity of the pion.
Analysis of our calculations shows that
$(\Sigma_{\pi^{+}}^{HT})^{0}$, $(\Sigma_{\pi^{+}}^{HT})^{res}$
higher-twist cross sections  are sensitive to the choice of the
infrared renormalon improved distribution amplitude of the pion.
We have also carried out comparative calculations in the
center-of-mass energy $\sqrt s=200\,\,GeV$ and obtained results
are displayed in Fig.7-Fig9. Analysis of our calculations at the
center-of-mass energies $\sqrt s=62.4\,\,GeV$ and $\sqrt
s=200\,\,GeV$, show that the increasing in the beam energy
contributions of higher twist effects to the cross section
decrease by about one-two order. In our calculations of the
higher-twist cross section of the process the dependence of the
transverse momentum of pion appears in the range of $(10^{-8} -
10^{-26})mb/GeV^2$. We think, that higher-twist cross section
obtained in this work should be observable at RHIC.

\section{Conclusions}\label{conc}
Proton-proton collisions are known to be the most elementary
interactions and form the very basis of our knowledge about the
nature of high energy collisions in general. Physicists, by and
large, hold the view quite firmly that the perturbative
quantum-chromodynamics  provides a general framework for the
studies on high energy particle-particle collisions. Obviously,
the unprecedented high energies attained at Large Hadron Collider
offer new window and opportunities to test the proposed QCD
dynamics with its pros and cons. However, we should remember that
LHC opens a new kinematical regime at high energy, where several
questions related to the description of the high-energy regime of
the QCD. Consequently, studies of proton-proton interactions at
the RHIC  and LHC could provide valuable  information on the QCD
dynamics at high energies. In this work the "naive
non-abelianization" (NNA) contributions of the higher-twist
Feynman diagrams to the large-$p_T$  inclusive pion production
cross section in proton-proton collisions are calculated. For
calculation of the higher-twist cross section the running coupling
constant approach is applied and infrared renormalon poles in the
cross section expression are revealed. Infrared renormalon induced
divergences is regularized by the means of the principal value
prescripton and the resummed expression (the Borel sum) for the
higher-twist cross section is find. It is abserved that, the
resummed higher-twist cross section differs  from that found using
the frozen coupling approach, in some regions, considerably. Also
we have demonstrated that higher-twist contributions to pion
production cross section in the proton-proton collisions have
important phenomenological consequences. Future RHIC and LHC
measurements will provide further tests of the dynamics of large
$p_T$ hadron production beyond leading twist.

\section*{Acknowledgments}
The work presented in this paper was started while one of the
authors, A.I.Ahmadov, was visiting the TH Division  of the CERN.
He would like to express his gratitude to the members of the TH
Division especially to the Prof. Jonathan Ellis  for their
hospitality. A.Ahmadov is also grateful to Stanley J.Brodsky for
useful discussions. Financial support by CERN is also gratefully
acknowledged.

\newpage

\begin{figure}[htb]
\vskip 0.002cm \epsfxsize 16cm \centerline{\epsfbox{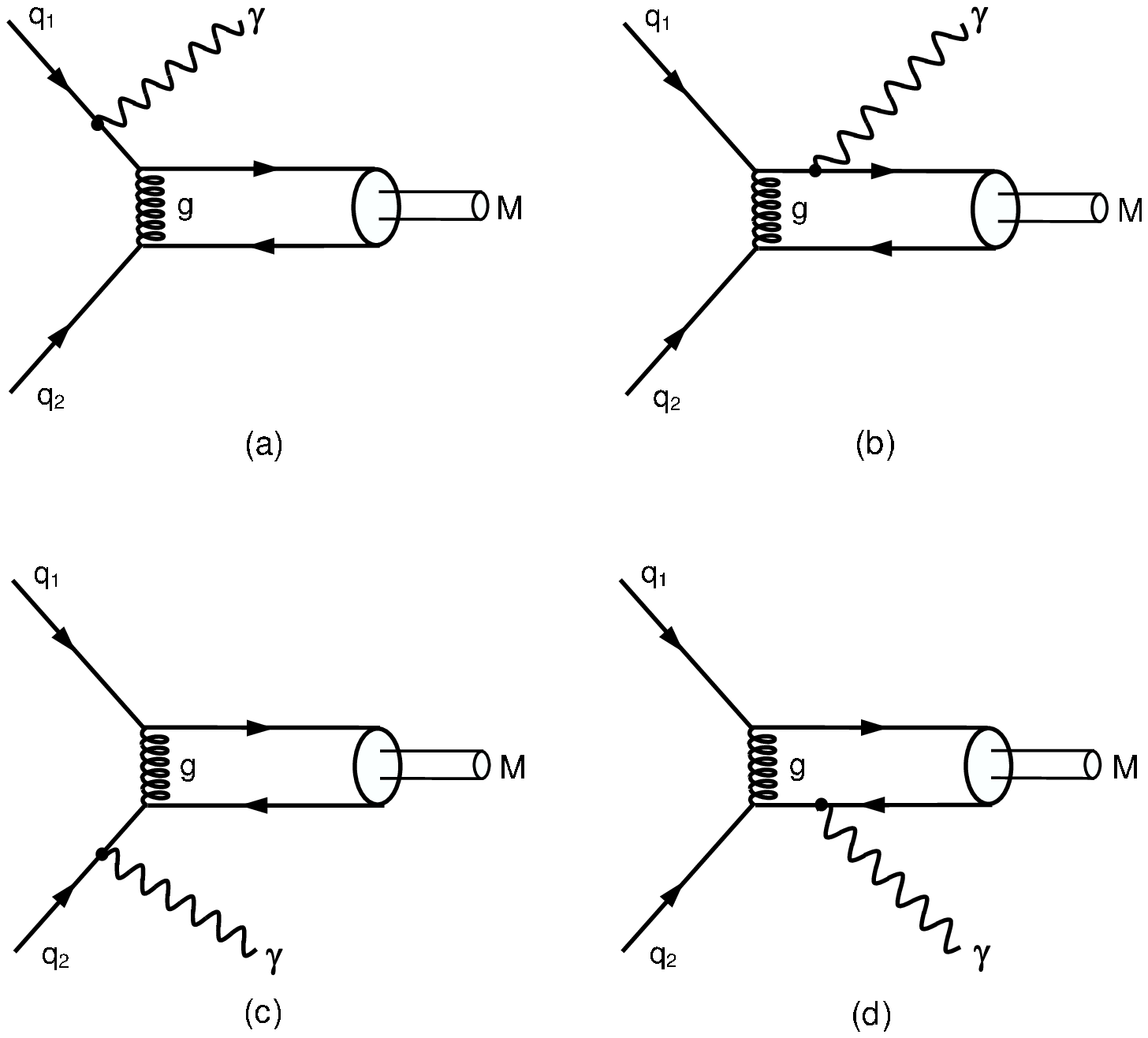}}
\vskip -7cm \caption{Feynman diagrams for the  higher-twist
subprocess, $q_1 q_2 \to \pi^{+}(or\,\,\pi^{-})\gamma.$}
\label{Fig1}
\end{figure}

\newpage

\begin{figure}[htb]
\vskip-1.2cm\epsfxsize 11.8cm \centerline{\epsfbox{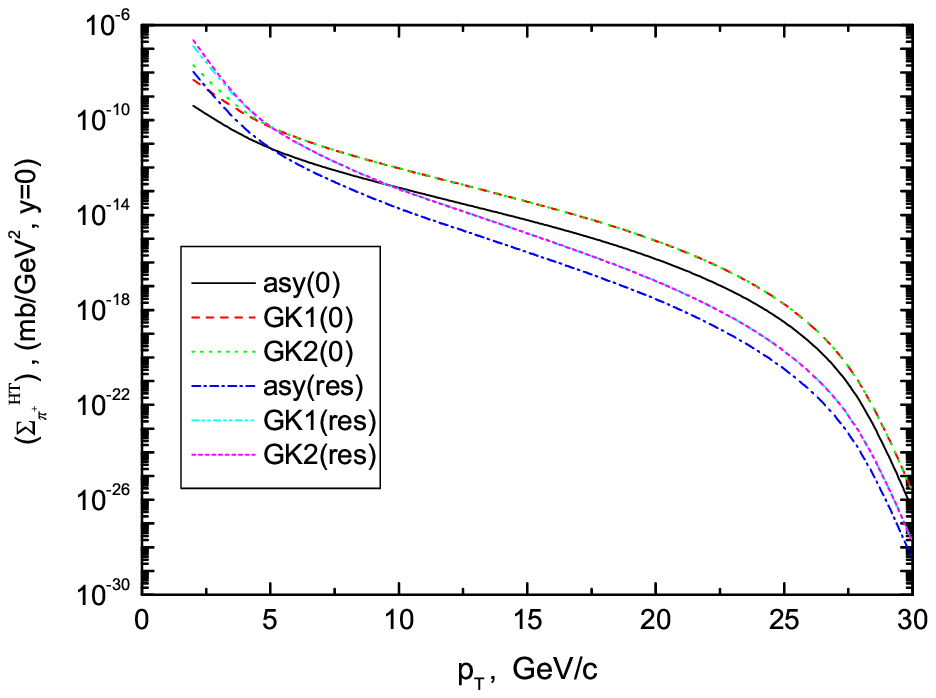}}
\vskip-0.2cm \caption{Higher-twist $\pi^{+}$ production cross
section $(\Sigma_{\pi^{+}}^{HT})$ as a function of the $p_{T}$
transverse momentum of the pion at the c.m.energy $\sqrt
s=62.4\,\, GeV$.}\label{Fig2}
 \vskip-1.0cm
\vskip 1.8cm \epsfxsize 11.8cm \centerline{\epsfbox{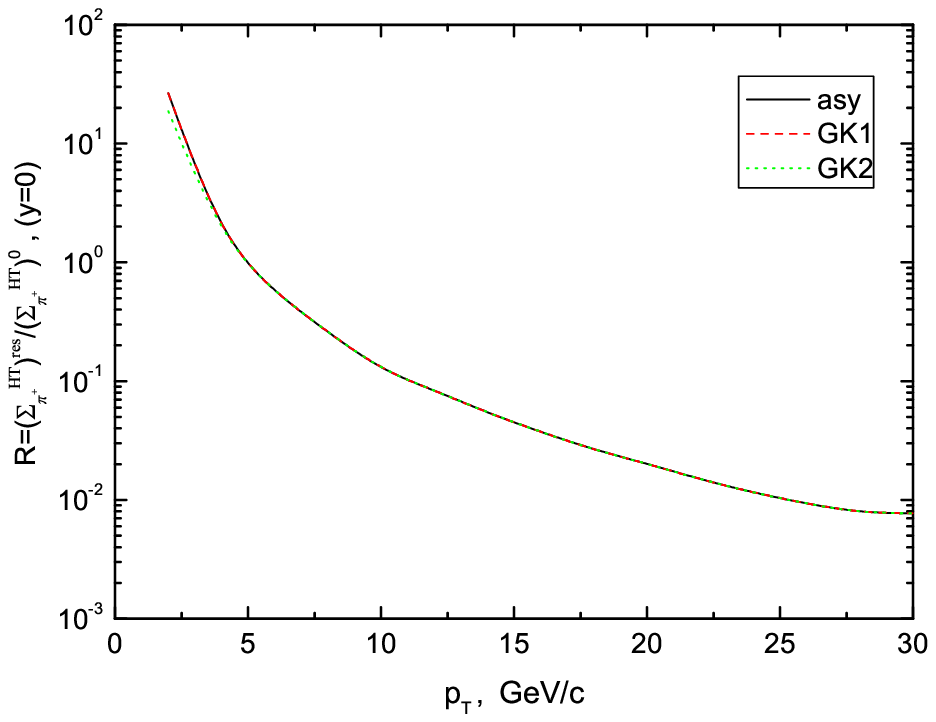}}
\vskip-0.05cm \caption{Ratio
$R=(\Sigma_{\pi^{+}}^{HT})^{res}/(\Sigma_{\pi^{+}}^{HT})^{0}$,
where higher-twist contribution are calculated for the pion
rapidity $y=0$ at the c.m.energy $\sqrt s=62.4\,\,GeV$ as a
function of the pion transverse momentum, $p_{T}$.}\label{Fig3}
\end{figure}

\newpage

\begin{figure}[htb]
 \vskip-1.2cm\epsfxsize 11.8cm \centerline{\epsfbox{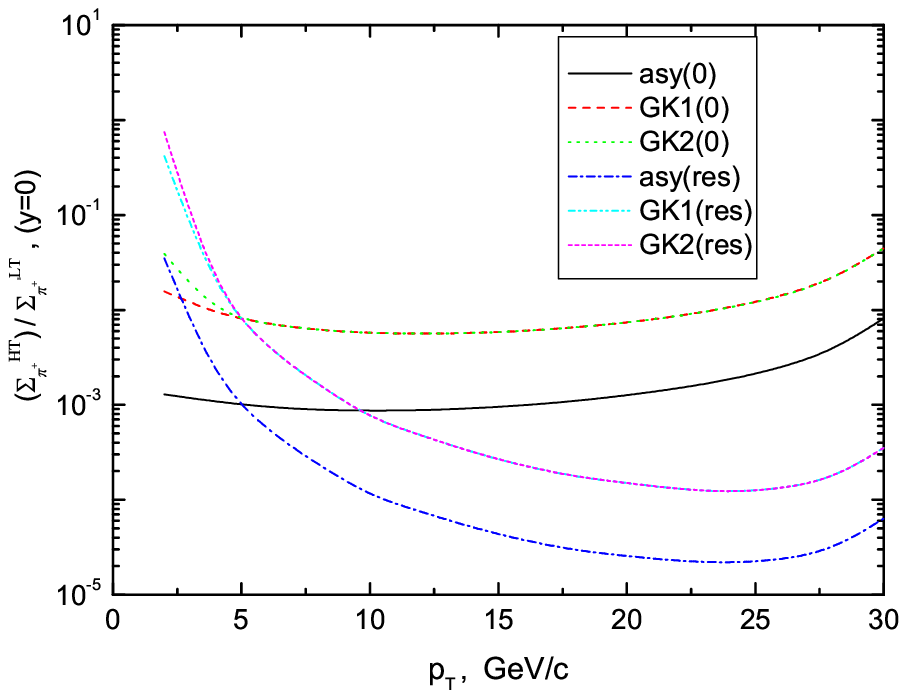}} \vskip-0.2cm
\caption{Ratio $(\Sigma_{\pi^{+}}^{HT})/\Sigma_{\pi^{+}}^{LT}$, as
a function of the $p_{T}$ transverse momentum of the pion at the
c.m.energy $\sqrt s=62.4\,\,GeV$.} \label{Fig4}
 \vskip 1.8cm
 \vskip-1.2cm\epsfxsize 11.8cm \centerline{\epsfbox{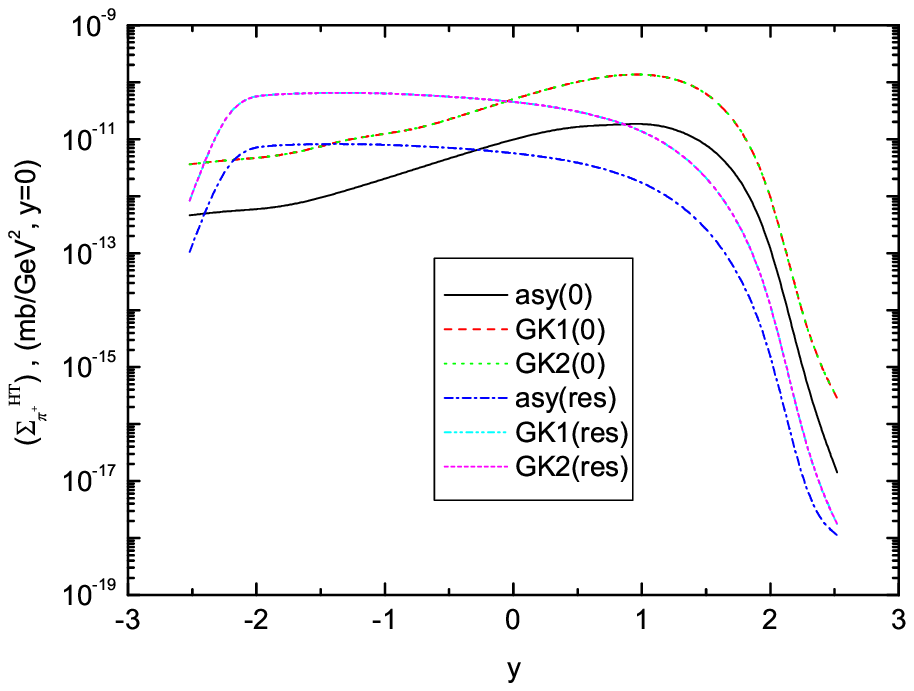}} \vskip-0.2cm
\caption{Higher-twist $\pi^{+}$ production cross section
$(\Sigma_{\pi^{+}}^{HT})$ , as a function of the $y$ rapidity of
the pion at the  transverse momentum of the pion $p_T=4.9\,\,
GeV/c$, at the c.m. energy $\sqrt s=62.4\,\, GeV$.} \label{Fig5}
\end{figure}

\newpage

\begin{figure}[htb]
\vskip-1.2cm \epsfxsize 11.8cm \centerline{\epsfbox{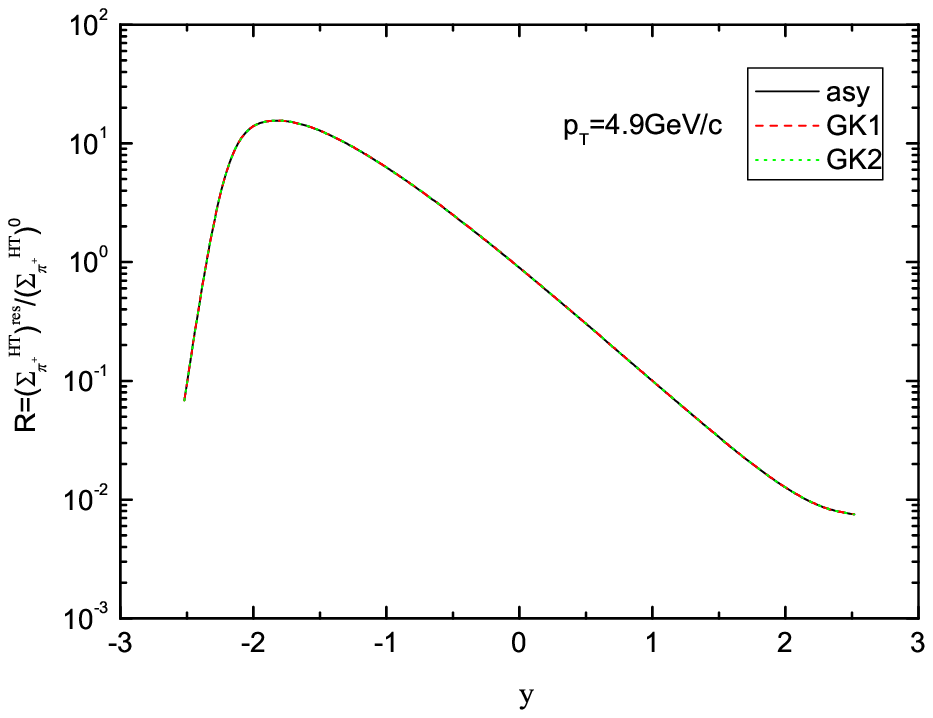}}
\vskip-0.2cm \caption{Ratio
$R=(\Sigma_{\pi^{+}}^{HT})^{res}/(\Sigma_{\pi^{+}}^{HT})^{0}$, as
a function of the $y$ rapidity of the pion at the  transverse
momentum of the pion $p_T=4.9\,\, GeV/c$, at the c.m. energy
$\sqrt s=62.4\,\, GeV$.} \label{Fig6}
\vskip-1.0cm
\vskip 1.8cm\epsfxsize 11.8cm \centerline{\epsfbox{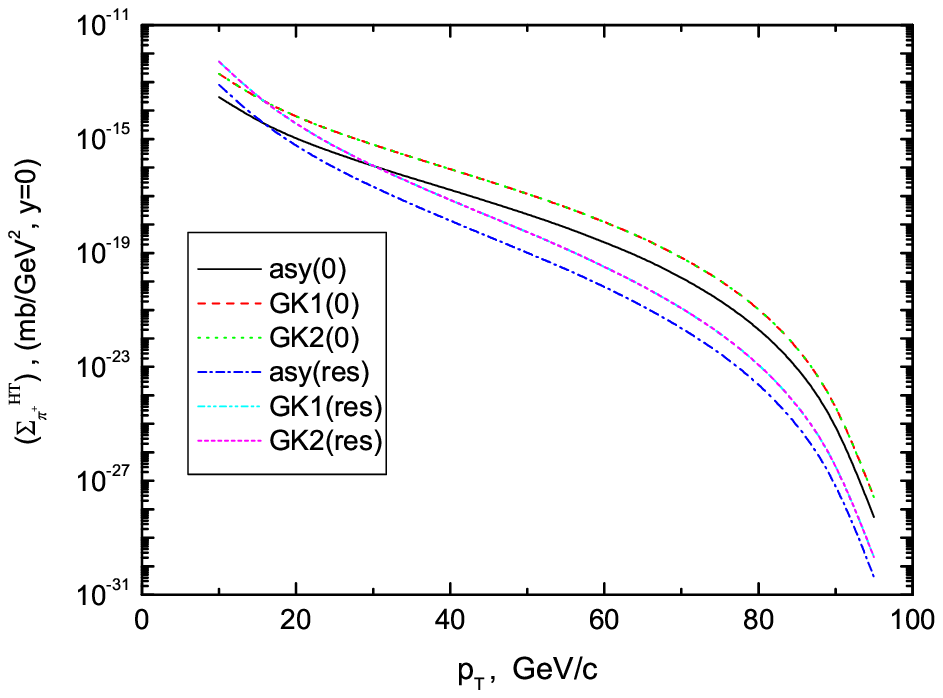}}
\vskip-0.05cm \caption{Higher-twist $\pi^{+}$ production cross
section $(\Sigma_{\pi^{+}}^{HT})$ as a function of the $p_{T}$
transverse momentum of the pion at the c.m.energy $\sqrt s=200\,\,
GeV$.} \label{Fig7}
\end{figure}

\newpage

\begin{figure}[htb]
\vskip-1.2cm \epsfxsize 11.8cm \centerline{\epsfbox{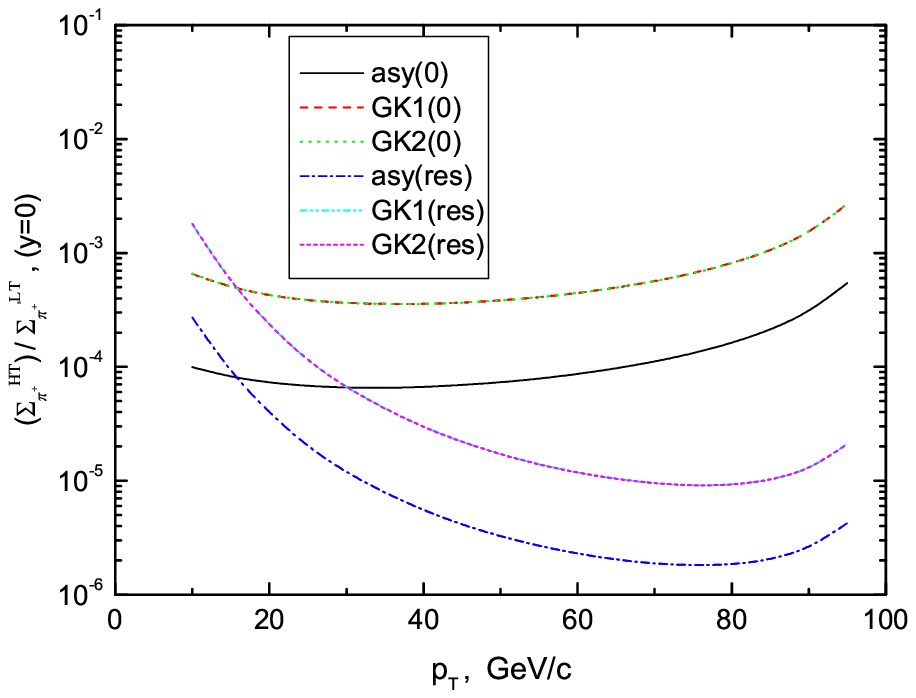}}
\vskip-0.2cm \caption{Ratio
$(\Sigma_{\pi^{+}}^{HT})/\Sigma_{\pi^{+}}^{LT}$, as a function of
the $p_{T}$ transverse momentum of the pion at the c.m.energy
$\sqrt s=200\,\,GeV$.} \label{Fig8}
\vskip-0.4cm
\vskip 0.8cm \epsfxsize 11.8cm \centerline{\epsfbox{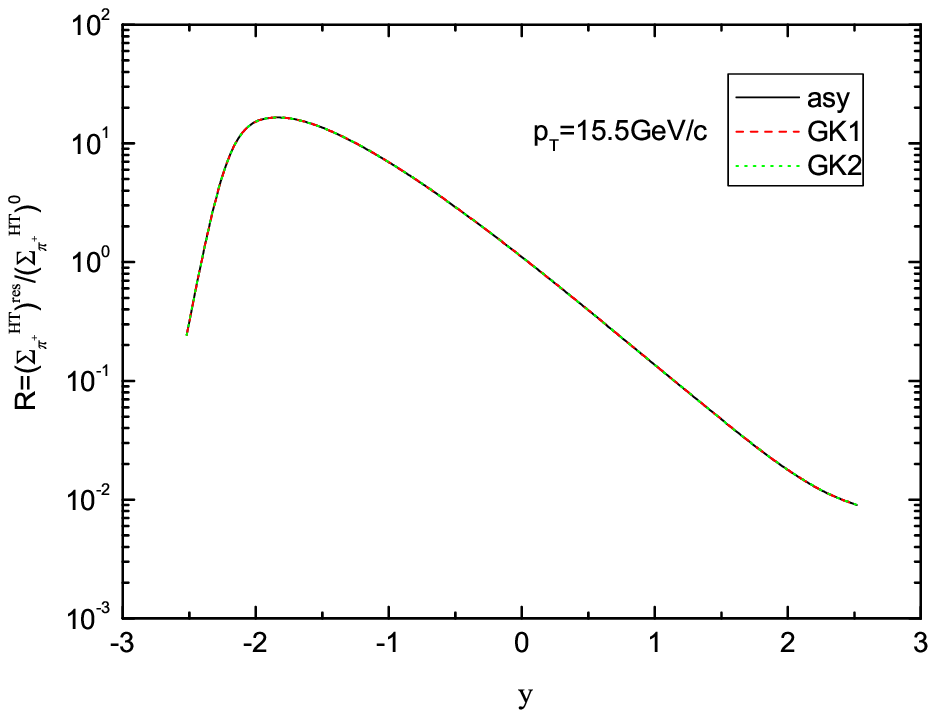}}
\vskip-0.2cm \caption{Ratio
$R=(\Sigma_{\pi^{+}}^{HT})^{res}/(\Sigma_{\pi^{+}}^{HT})^{0}$, as
a function of the $y$ rapidity of the pion at the  transverse
momentum of the pion $p_T=15.5\,\, GeV/c$, at the c.m. energy
$\sqrt s=200\,\, GeV$.} \label{Fig9}
\end{figure}

\end{document}